# Cancer Dissemination: A Consequence of limited Carrying Capacity?

**Thomas S. Deisboeck** [*] **and Zhihui Wang**

Complex Biosystems Modeling Laboratory, Harvard-MIT (HST) Athinoula A. Martinos Center for Biomedical Imaging, Massachusetts General Hospital, Charlestown, MA 02129, USA.

**[*]Corresponding Author:**

Thomas S. Deisboeck, M.D.
Complex Biosystems Modeling Laboratory
Harvard-MIT (HST) Athinoula A. Martinos Center for Biomedical Imaging
Massachusetts General Hospital-East, 2301
Bldg. 149, 13th Street
Charlestown, MA 02129
Tel: 617-724-1845
Fax: 617-726-7422
Email: deisboec@helix.mgh.harvard.edu

## ABSTRACT

Assuming that there is feedback between an expanding cancer system and its organ-typical microenvironment, we argue here that such local tumor growth is initially guided by *co-existence* rather than competition with the surrounding tissue. We then present a novel concept that understands cancer dissemination as a biological mechanism to evade the specific *carrying capacity* limit of its host organ. This conceptual framework allows us to relate the tumor system's volumetric growth rate to the host organ's functionality-conveying composite infrastructure, and, intriguingly, already provides useful insights into several clinical findings.

**ARTICLE**

Metastasis is a common complication of many solid cancer types and generally indicates advanced stage disease. What, however, varies considerably from patient to patient are for instance dynamics and extent of the dissemination. To our knowledge, up until now the process of metastasis has been primarily investigated from an experimental tumor biology perspective, describing it as an intricately complicated process that involves multiple steps including cell detachment, cell-matrix interaction, tissue infiltration and angiogenesis [1–4]. While these works undoubtedly led to significant insights over the years, a detailed understanding as to the general dynamics driving the *onset* of metastasis has yet to emerge.

It is generally accepted that tumors crucially depend on extrinsic nutrients provided by the host [e.g., 5–7] in conjunction with their auto- and paracrine produced growth factors [8]. Here, we therefore argue that, for a tumor, 'success', with regards to its overall growth would be ill defined as merely gaining a *competitive* advantage over a rapidly failing host organ. Rather, characterizing the underlying relationship as *coexisting*, evolutionary success should be assessed by as to how well this tumor manages its interaction with the host site so that *continuous* malignant growth is ensured within a well nourishing since persevering host organ. Following this novel conceptual framework, metastasis should be triggered when the *carrying capacity* of the primary host organ (both in terms of biochemistry and biomechanics) is about to be exceeded. While the biological processes involved in cancer expansion may lead to some upward adjustment of the carrying capacity, presumably these parenchymal and stromal compensation mechanisms are limited, and tumor cells will eventually spread to distant sites. We note that the so-called 'carrying capacity' of a given environment has been widely studied in the area of ecology to estimate human population dynamics [9,10]. Moreover, as a parameter, carrying capacity has already been implemented in several mathematical models to study the tumor's adaptive responses [11,12], but the idea in these works was still constrained by the point that the growth of tumor cell populations is due to competition for limited resources.

We argue that malignant cancer may not be geared towards conquering the host that it so critically depends on but, particularly in its *early* stages, benefits from maintaining a state of co-existence. Since there is no apparent benefit for the hosting organ, this 'co-existing' relationship is profoundly one-sided and, for the case of malignant tumors, overall *finite*. Ultimately, the increasingly aggressive makeup of the cancerous cells overcomes the limited biomechanical and -chemical compensation mechanisms available to the original site, then triggering metastasis as a mere 'escape' mechanism in a futile effort to avoid the inevitable. That is, while such dissemination to secondary sites is, according to this new concept, meant to prolong the overall coexistence, the mounting damage that it causes to the patient's delicate physiology makes the disease less controllable and thus usually worsens the overall prognosis [e.g., 13]. We conjecture that cancer may in fact not fit the usual characterization of an *unregulated* lopsided growth process where dissemination is primarily means of expansive 'colonialism', but behaves as a coexisting growth process where invasion and metastasis serve as tools to *evade* the detrimental effects of growing microenvironmental constraints. While all this at first may seem rather odd, our concept merely argues for continuous *feedback* between tumor and microenvironment, and, within limits, the possibility of dynamic *adaptation* on both sides. We note that the experimental evidence supporting such feedback is mounting [e.g., for a recent review on melanoma-microenvironment interaction see 14]. Following our hypothesis, the *optimization* goal of the tumor system is *maximizing* its spatio-temporal expansion rate[1] while *maintaining* the nourishing microenvironmental conditions *as much as necessary and as long as possible* to 'selfishly' sustain this maximum growth rate.

Moving this qualitative concept now into a more theoretical framework, the carrying capacity, $C_C$, of a given host organ denotes the maximum tumor volume, $V_{Tum}$ that can be sustained by that tissue's composite volume infrastructure[2], $V_{Tis}$, without causing organ malfunction. To properly reflect this function-volume relationship, we start with defining $C_C$ as

$$C_C = \frac{1}{\left(\dfrac{F_{Tis}}{V_{Tis}}\right)} \tag{1}$$

---

[1] Employing proteolysis, invasion, adhesion, angiogenesis and related epi/genetic progression.
[2] Comprising biomechanical and biochemical components.

where $F_{Tis}$ represents the level of specialization or functionality of a tissue (depending on both tissue and the parameter (set) used to reflect its functionality). While $V_{Tis}$ and $C_C$ are being defined in volume units, $F_{Tis}$ is defined as a non-dimensional parameter for now. We argue here that the extent of functional connectivity if not circuitry in 'evolved' and thus highly differentiated and specialized tissues ensures that $F_{Tis}$ equals and more likely exceeds the value of its corresponding unit of tissue volume and thus $C_C \leq 1$. Now, since, at least during the early tumor growth phase,

$$V_{Tum} \leq C_C \tag{2}$$

it follows conceptually that the *critical* threshold for $V_{Tum}$ to trigger the onset of metastasis should be $\leq 1$. Eq. (2) indicates that the higher the tissue's function-volume relationship, i.e. the more differentiated the host organ, the smaller its $C_C$ and thus the *smaller* the maximum tumor volume that this particular host organ can sustain. We note that on-site tumor dissemination or invasion involves proteolysis, i.e., enzymatic degradation of adjacent parenchyma. We hypothesize that this disruptive local expansion sufficiently damages the tissue's infrastructure and consequently, impacts and ultimately impairs the organ's functionality. According to Eq. (1), any such reduction in $F_{Tis}$ will lead to an *increase* in primary site $C_C$, much like any increase in $V_{Tis}$ through, for instance, tumor-induced angiogenesis [15][3].

One may further argue that cancer *dissemination* towards multiple organ sites *i, j ... n* must be guided by the premise of a *more permissive* $C_{C\ [Total]}$. However, following a *complex systems* concept[4] multi-organ site physiology should constitute a larger increase in $F_{[Total]}$ than in $V_{[Total]}$ and, in its non-diseased state, thus overall yield a *reduced* $C_{C\ [Total]}$. Therefore, in order to generate any increase in $C_{C\ [Total]}$ the process of metastasis would have to cause a marked damage to multi-organ functionality early on. This should result in a sizeable selection pressure that fuels tumor *progression* on a systemic level and that expands $V_{Tum\ [Total]}$, worsens the patient's prognosis and thus reflects the common clinical scenario. That said, the volumetric growth *rate* should however be a more relevant indicator for the impact the cancer systems has, than a given volume alone and so we propose that

$$\frac{\Delta V_{Tum}}{\Delta t} \leq \frac{\Delta C_C}{\Delta t} \qquad (2.1)$$

---

[3] For the case of angiogenesis, this tumor-organ *coexistence* can be characterized as 'commensalisms', as only one side, the tumor, benefits from this form of *cooperation*. In this context, for a recent article on the application of 'game theory' to cancer growth see [16], and for theoretical modeling approaches see e.g. [17] and [18].
[4] The understanding, that a system's emergent behavior *exceeds* the mere sum of its component parts. For more information on this topic see e.g. [19] and references therein.

Eq. (2.1) states that gaining modest volume over a much larger timeframe is less likely to threaten an organ's carrying capacity, whereas a rapid change even of a relative small tumor volume can quickly approximate a set, limited carrying capacity. However, in reality, $C_C$ is not static either and thus Eq. (2.1) accounts for dynamic (biomechanical and biochemical) tissue compensation mechanisms, that require sufficient time to adjust properly and overall are finite. Taken together, we deduce from Eqs. (1) and (2.1) that *aggressive tumor growth within a highly differentiated organ causes early onset of symptoms.* That is, according to Eq. (1) $C_C$ should be rather low to begin with in most mammalian organs, and thus the cancer growth induced $\Delta V_{Tis}$ (= *increase* in composite tissue infrastructure volume) and/or $\Delta F_{Tis}$ (= *reduction* in functionality) has to be substantial to expand $C_C$ sufficiently, and do so rather quickly, in an effort to delay early onset of metastasis. This, in turn, supports the notion that a tumor progresses locally first, *prior* to any metastasis. At *later* stages, once dissemination occurs, the now systemic disease rapidly becomes even more aggressive, i.e. rendering it 'de facto' *competitive*, for the reasons detailed above. Taken together, we argue that it is this limit in carrying capacity, both locally and globally, which *drives* cancer system progression *and* expansion and that therefore ultimately threatens the state of coexistence the tumor so critical depends on.

We note that the left term in Eq. (2.1), i.e. $\Delta V_{Tum}/\Delta t$, is not restricted to a particular tumor growth model and can follow e.g. logistic, Gompertz or Universal scaling laws [20,21]. Rewriting Eq. (1) as $C_C(t) = \frac{V_{Tis}(t)}{F_{Tis}(t)}$, the rate of change of $C_C$ becomes

$$\frac{d}{dt}(Cc(t)) = k_1(t)\frac{dV_{Tis}(t)}{dt} - k_2(t)\,V_{Tis}(t)\frac{dF_{Tis}(t)}{dt} \qquad (3)$$

where the variables $k_1(t) = 1/F_{Tis}(t)$ and $k_2(t) = 1/F_{Tis}^2(t)$. As $F_{Tis}(t) > 0$ and $V_{Tis}(t) > 0$, $k_1(t) > 0$ and $k_2(t) > 0$. For simplification, $V_{Tis}(t)$ is treated in a generic way, i.e. with a unit volume of 1, regardless of its real metric volume. Since we argue that $V_{Tis}(t)$ is *equal to or smaller than* $F_{Tis}(t)$, it follows that $F_{Tis}(t) > 1$. Thus, the relation between $k_1(t)$ and $k_2(t)$ is $k_2(t) \leq k_1(t)$. From this, we can deduce that the impact a temporary change in $V_{Tis}$ has on increasing $C_C$ is *greater than or equal* to that caused by dynamic variations in $F_{Tis}$. Combining Eqs. (2.1) and (3), we deduce that a *sufficient* condition, but not a necessary and sufficient condition, for continued tumor growth is

$$\frac{d}{dt}(V_{Tum}(t)) \leq k_1(t)\frac{dV_{Tis}(t)}{dt} - k_2(t)V_{Tis}(t)\frac{dF_{Tis}(t)}{dt} \quad (4)$$

Eq. (4) summarizes the dynamic relationships between changes in tumor volume, host tissue functionality and its composite infrastructure. Specifically, it states that if the growth rate of the tumor is less than or equal to the adjustment rate of $C_C$, the host organ's environmental setting remains permissive for *on site* cancer growth. Consequently, we hypothesize that once the tumor's growth rate exceeds the compensation mechanisms available to the tissue, more metastatic phenotypes will be selected within the heterogeneous tumor cell population.

This new theoretical framework offers an intriguing opportunity to conceptualize several scenarios that have significance for the clinical situation.

- For instance, the finding that a tumor-induced increase in the rate of change of $V_{Tis}$ has likely a more substantial impact on moderating the primary or host organ's $C_C$ than a change rate reduction in its tissue's function can achieve, may suggest that the tumor, as an opportunistic system, employs *neo*-infrastructure building processes (and here most notably neo-vascularization) *first and more so* than to operate with infiltrative tissue destruction. This then does seem to support our argument of the tumor system initially striving for coexistence rather than competition with its nourishing mircoenvironment. We note, however, that in increasing $V_{Tis}$, the process of angiogenesis involves endothelial cell migration towards the chemoattractant secreting tumor. The latter should impact, possibly damage the native parenchymal infrastructure (in addition to the damage done by the advancing tumor cells); hence, by means of reducing $F_{Tis}$ tumor, angiogenesis may lead to an even more substantial increase in $C_C$. Following this line of thought, we conjecture that, in an effort to achieving local control, targeting tumor angiogenesis should be more promising clinically than trying to reduce the activity of the tumor's infiltration-mediating enzymes, i.e. proteases. And indeed, recent clinical studies showed no improvement of outcome for glioma patients treated with the metalloproteinase-inhibitor marimastat [22] whereas studies with anti-angiogenetic drugs already demonstrated significant clinical potential [23,24].

- Secondly, if indeed a high level of functionality per unit volume of tissue is the result of evolutionary differentiation, it is reasonable to argue that with increasing age, the rate of decline in a tissue's highly specialized functionality, $F_{Tis}$, exceeds the decay in its composite tissue infrastructure, $V_{Tis}$. According to Eq. (1), the result would be an aging-related *increase* in the tissues' $C_C$. Interestingly this scenario could explain why in the elderly population the incidence of cancer increases while concomitantly, the process of metastasis appears to be slowed down [25–27]. In turn, in younger age patients the tissues' carrying capacity should be relatively low on average and as such tumor dissemination would generally start earlier and move faster. Intriguingly, this argument is again already supported by clinical findings for younger breast cancer patients (< age 35 years of age) where aggressive histopathological features, including the number of metastatic lymph nodes, relate to significantly higher probability of relapse and thus an overall lower 5-year survival [28].

- Lastly, following the same line of thought, any surgical debulking with removal of tumor-harboring tissue structure may create an even larger functional damage and as such yield an increased carrying capacity for the organ residue. This presents the following dilemma: while a 'total' resection with tumor-free margins in the healthy tissue is common goal in clinics [e.g., 29], following our conjecture, therapeutic intervention geared towards aggressively reducing tumor burden can paradoxically render the remaining, damaged 'soil' [30] *more permissive* for any tumor cells left behind and thus may *facilitate* recurrence.

Admittedly, advancing this theoretical framework into a clinically useful, *quantitative* method seems to pose several formidable challenges at first, such as assessing 'structure-function' relationships *per organ and per patient*. However, on a second thought, a multitude of quantitative tests is already used in clinical practice including e.g. creatinine clearance tests to assess kidney function, lung spirometry tests and enzymatic tests to quantify liver function and one can extrapolate, that continuous improvements of available *in vivo* imaging modalities [31] will allow assessing ever smaller structural entities.

We conclude that if our hypothesis holds true, one may have to accept yet arguably should also be able to readily exploit that in clinics success in battling advanced-stage disease means *managing co-existence* with the tumor system more so than detect, target and eradicate each and every cancer cell in the patient's body. This innovative therapeutic approach would focus on patient-specific assessment and monitoring of the diseased tissues' dynamically changing carrying capacity in an effort to clinically manage its change rate and thus attempt controlling tumor expansion. Given the significant impact this would have on oncology-related health care, experimental *in vitro* and *in silico* studies are warranted and necessary as a first step to test this intriguing theoretical framework thoroughly.

## ACKNOWLEDGEMENTS

This work has been supported in part by NIH grants CA 085139 and CA 113004 and by the Harvard-MIT (HST) Athinoula A. Martinos Center for Biomedical Imaging and the Department of Radiology at Massachusetts General Hospital. We thank Dr. Caterina Guiot (Dip. Neuroscience-INFM, Università di Torino) and Dr. Le Zhang (Complex Biosystems Modeling Laboratory, Massachusetts General Hospital) for critical review of the manuscript.

Deisboeck, T.S. & Wang, Z.: *Cancer Dissemination*[2] Bogenrieder T, Herlyn M. Axis of evil: molecular mechanisms of cancer metastasis. Oncogene 2003;22(42):6524–6536.

[3] Bernards R, Weinberg RA. A progression puzzle. Nature 2002;418(6900):823–824.

[4] Friedl P, Wolf K. Tumour-cell invasion and migration: diversity and escape mechanisms. Nat Rev Cancer 2003;3(5):362–374.

[5] Folkman J. Tumor angiogenesis: therapeutic implications. N Engl J Med 1971;285(21):1182–1186.

[6] Carmeliet P. Angiogenesis in life, disease and medicine. Nature 2005;438(7070):932–936.

[7] Jain RK. Normalization of tumor vasculature: an emerging concept in antiangiogenic therapy. Science 2005;307(5706):58–62.

[8] Nicolson GL. Cancer progression and growth: relationship of paracrine and autocrine growth mechanisms to organ preference of metastasis. Exp Cell Res 1993;204(2):171–80.

[9] Cohen JE. Population growth and earth's human carrying capacity. Science 1995;269:341–346.

[10] Meyer PS, Ausubel JH. Carrying capacity: a model with logistically varying limits. Technol Forecast Soc Change 1999;61(3):209–214.

[11] Michelson S, Leith JT. Growth factors and growth control of heterogeneous cell populations. Bull Math Biol 1993;55(5):993–1011.

[12] Michelson S, Leith JT. Positive feedback and angiogenesis in tumor growth control. Bull Math Biol 1997;59(2):233–254.

[13] Myrdal G, Lambe M, Gustafsson G, Nilsson K, Stahle E. Survival in primary lung cancer potentially cured by operation: influence of tumor stage and clinical characteristics. Ann Thorac Surg 2003;75(2):356–363.

[14] Postovit LM, Seftor EA, Seftor RE, Hendrix MJC. Influence of the microenvironment on melanoma cell fate determination and phenotype. Cancer Res 2006;66(16):7833–7836.

[15] Carmeliet P, Jain RK. Angiogenesis in cancer and other diseases. Nature 2000;407(6801):249–257.

[16] Axelrod R, Axlerod DE, Pienta KJ. Evolution of cooperation among tumor cells. Proc Natl Acad Sci USA 2006;103 (36):13474–13479.

[17] Gatenby RA, Vincent TL. An evolutionary model of carcinogenesis. Cancer Res 2003; 63(19): 6212-6220.
9